\documentstyle[11pt,newpasp,twoside,psfig,epsfig]{article}
\def\edcomment#1{\iffalse\marginpar{\raggedright\sl#1\/}\else\relax\fi}
\reversemarginpar

\begin{document}
\title{An Unbiased Survey for Outflows in the W3 and W5 Star-Formation Regions
}
 \author{Derek E. Bretherton \& Toby J.T. Moore}
\affil{Astrophysics Research Institute, Liverpool John Moores University, 
Twelve Quays House, Egerton Wharf, Birkenhead, CH41 1LD, UK
}
\author{Naomi A. Ridge}
\affil{FCRAO, Department of Astronomy, University of Massachussetts, Amherst, MA 01003, USA
}


\section{Introduction}

During their birth all stars undergo periods of copious mass loss,
frequently characterized by the occurrence of bipolar outflows. These
outflows are believed to play a fundamental role in the star formation
process. However the exact outflow generating method is obscure at
present. To elucidate this problem we are investigating whether the
flow properties are correlated over the entire protostellar mass
spectrum.

Progress in this area requires that we assemble a statistically valid
sample of high-mass outflow systems. This is necessary since existing
catalogues of such objects are heterogeneous and statistically
incomplete.

\section{Target Selection and Observations}

In order to produce a representative outflow sample a new $^{12}$CO
survey of the W3 and W5 molecular clouds was initiated at the 14-metre
Five College Radio Astronomy Observatory (FCRAO) to look for high
velocity gas. W3/5 are intermediate- to high-mass star-forming regions
located at a distance of 2 kpc in the Perseus arm.

Survey observations were carried out during April 1999, April 2000,
and November 2000. Position-switched $^{12}$CO 1--0 emission mapping
observations were obtained with beam-width (44$''$ at 115\,GHz)
sampling, using the SEQUOIA spectral plane array.

\section{Preliminary Results}

Since the youngest sources may emit insufficiently in the infrared to
be detected by IRAS, the whole data-set was examined for high-velocity
gas, indicative of molecular outflows, {\it independent} of IRAS data.

Due to the large volume of data an automated potential outflow
detection routine was developed.  Potential outflow candidates were
identified on the basis of line wings present in their spectra. A
Gaussian profile was fitted to each spectrum, and then subtracted to
leave a residual. The presence of a line wing manifests itself as
excess emission (above the noise level) in the residual after
subtraction of the Gaussian. (For a more detailed discussion of the
potential outflow detection routine see the forthcoming series of papers by
Ridge, Bretherton \& Moore and Bretherton, Ridge \& Moore).

The outflow detection routine successfully identified all known
outflows in W3 and W5 (W3-IRS5, W3 (OH), IC1805-W, AFGL 4029 \& AFGL
437). In addition it flagged $\sim$40 regions which might harbour
molecular outflows.

Follow-up Nyquist sampled observations of 24 of these outflow
candidates were carried out at FCRAO during Spring
2001. Position-switched $^{13}$CO observations were also obtained at the
central position in order to derive the $^{12}$CO optical depth, and to
check for multiple-velocity cloud components, which might confuse the
outflow detection algorithm. Seven of the potential outflows were
discarded on the basis of the $^{13}$CO observations.

\begin{figure}[ht!]
\centering
\vspace*{4.6cm}
   \leavevmode \includegraphics{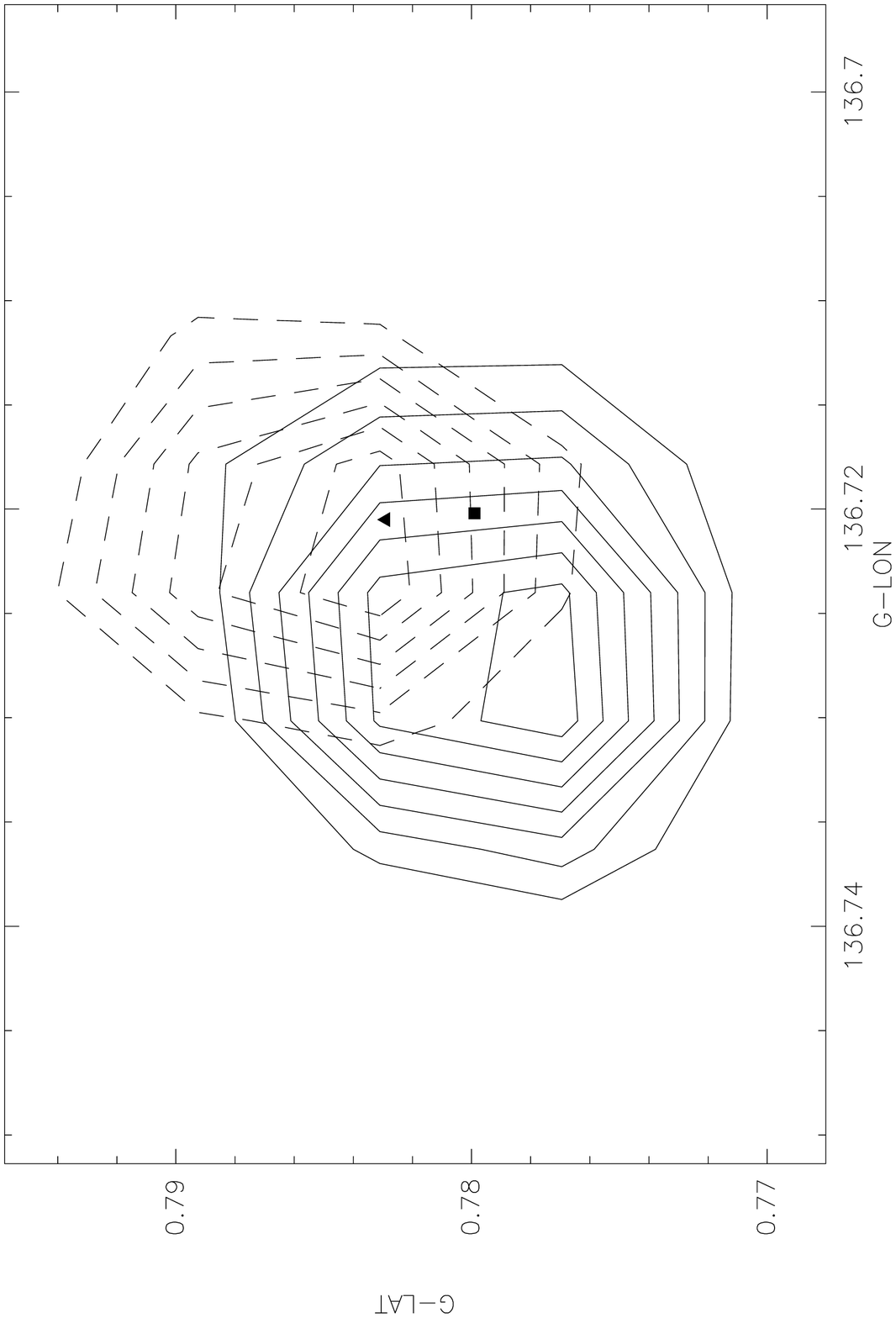}
\end{figure}
{\bf Figure 1:} Contour map of the integrated $^{12}$CO 1--0 emission
from a new outflow source in W5. (This source was confirmed as a {\it
bona fide} outflow by observations at the Nobeyama 45m Radio
Telescope). This fully-sampled map was made at the FCRAO during the
follow-up observations. Spectra corresponding to this region were
flagged by the outflow detection routine during reduction of the
preliminary programme data. The solid-line contours show the red-lobe
emission integrated between -40.8 and -35.8 kms$^{-1}$. Dashed
contours show the blue-lobe emission integrated between -50.8 and
-45.8 kms$^{-1}$.  Base contour is 1.8 K kms$^{-1}$ with contour
intervals of 0.5 K kms$^{-1}$. The triangle indicates the position of
an IRAS source, whilst the square indicates the presence of a MSX
source.

\section{Further Work}

The next stage of the programme (currently in progress) is the
acquisition of high-resolution maps of those outflow candidates not
rejected by the FCRAO follow-up observations. High-resolution maps
will confirm whether the targets are {\it bona-fide} outflows.

Once we have assembled a statistically significant outflow
sample we will be able to comment on the applicability of current
theories of outflow generating mechanisms. Furthermore the variation
of mass and momentum in outflows with source luminosity will be
examined, thus enabling stronger constraints to be placed on the
actual physics of specific flow models.

\end{document}